\documentclass[aps,prd,onecolumn,nofootinbib,notitlepage,11pt]{revtex4-2}
\usepackage{amsmath,amssymb,amstext,graphicx,color,xfrac,braket,accents,tabulary,booktabs,xcolor,bbold,hyperref,natbib,setspace}
\begin{document}
\thispagestyle{empty}

\newcommand{\be}{\begin{equation}}
\newcommand{\ee}{\end{equation}}
\newcommand{\bea}{\begin{eqnarray}}
\newcommand{\eea}{\end{eqnarray}}
\newcommand{\hst}{\widetilde{\mathcal{H}}} 
\newcommand{\iso}{\dot{=}}
\newcommand{\Dim}{\textrm{dim\,}}
\newcommand{\Tr}{\textrm{Tr\,}}
\newcommand{\hs}{\mathcal{H}} 
\newcommand{\E}{\widetilde{E}}
\newcommand{\thetat}{\tilde{\theta}}
\newcommand{\ham}{\widehat{H}}
\newcommand{\intham}{\widehat{H}_{\rm{int}}}
\newcommand{\selfham}{\widehat{H}_{\rm{self}}}
\newcommand{\dt}{\delta_t}
\newcommand{\trace}{\mathrm{Tr}}
\def\bra#1{\langle #1\rvert}
\def\ket#1{\lvert #1\rangle}
\newcommand{\draftnote}[1]{\textbf{\color{red}[#1]}}
\def\andname{\unskip,}

\baselineskip=14pt

\title{Toward a Phenomenologically Acceptable Quantum Cyclic Universe}
\author{Sean M. Carroll$^{abc}$}
\email{seancarroll@gmail.com}
\author{Nadiia Diachenko$^a$}
\email{ndiache1@jhu.edu}
\author{Saakshi Dulani$^{ab}$}
\email{saakshi.dulani@jhu.edu}
\affiliation{$^a$Department of Physics \& Astronomy, Johns Hopkins University, Baltimore MD 21218,}
\affiliation{$^b$Department of Philosophy, Johns Hopkins University, Baltimore MD 21218}
\affiliation{$^c$Santa Fe Institute, Santa Fe, NM 87501} 

\begin{abstract}
We put forward a quantum model of cosmology that is exactly periodic but avoids the Boltzmann Brain problem.
If the universe is described by a quantum state evolving unitarily in a finite-dimensional Hilbert space, its evolution will be recurrent: given enough time, the state will return arbitrarily close to its initial state.
There is a worry that such a scenario cannot be phenomenologically acceptable, because the state will spend most of its time in a high-entropy equilibrium macrostate, with rare fluctuations downward in entropy, and the vast majority of observers will be minimal fluctuations away from equilibrium, or ``Boltzmann Brains."
Here we show that this is not necessarily true.
If the differences in energy eigenvalues are commensurable, the evolution is not simple recurrent, but exactly periodic.
Moreover, if the state starts at minimum thermodynamic entropy, its evolution can feature a distinguished entropy excursion that is much more pronounced than one would expect from the conventional expression $P(\Delta S) \propto \exp(-\Delta S)$.
This excursion could represent our Big Bang, with relatively few Boltzmann fluctuations occurring in the subsequent equilibrium phase before a Big Crunch occurs and the cycle begins again.
We speculate on the spacetime interpretation of this kind of quantum universe.
\end{abstract}

\maketitle

\noindent``The eternal hourglass of existence is turned over again and again, and you with it, speck of dust!" 
\makebox[\textwidth][r]{-- F. Nietzsche \cite{nietzsche1974gay}}

\section{Introduction}\label{intro}

Our observable universe is manifestly asymmetric in time.
It began approximately 14 billion years ago in a hot Big Bang, and toward the future it is becoming emptier and colder; most stars that will ever form have already formed \cite{Madau:2014bja}.
If the vacuum energy is an unchanging cosmological constant, which is plausible but not certain, the universe will continue to expand and asymptotically enter a de~Sitter phase, in accordance with the Cosmic No-Hair Theorem \cite{Wald:1983ky,Barrow:1989wp,Carroll:2017kjo}. 

This state of affairs suggests a number of questions.
Why is there a temporal imbalance at all?
More specifically, conditions near the Big Bang were very low-entropy, which is responsible for the thermodynamic arrow of time \cite{penrose1979singularities,albert2003time,carroll2010eternity}.
This is anthropically fortunate (as life can only exist far from equilibrium), but physically seems fine-tuned, as the number of low-entropy states is much smaller than the number of high-entropy ones; what is the origin of this low-entropy boundary condition?
Why is the vacuum energy, which from the perspective of effective field theory could naturally be near the Planck scale, so small but not precisely zero?
And what happened at the very beginning, where classical general relativity predicts a singularity?

Most cosmological scenarios take this temporal asymmetry as given, but some attempt to explain it, or at least place it in a larger context.
One possibility is that the universe will eventually recollapse, and entropy will decrease toward a future Big Crunch \cite{gold1962arrow,Price:1993hr,kiefer1995arrow}.
Hawking once suggested that such a behavior was natural in quantum cosmology \cite{hawking1985arrow}, but that idea was later refuted \cite{page1985will,hawking1993origin}.
At face value, such models are temporally even-handed, but invoking unnaturally low-entropy conditions in both future and past isn't an obvious improvement over simply invoking them in the past.
Another globally time-symmetric alternative is to have entropy increasing asymptotically in both the far past and the future; this scenario requires the existence of a time-reversed pre-Big-Bang era of some kind \cite{aguirre2003inflation,Carroll:2004pn,hartle2012arrows,Barbour:2013jya,Goldstein:2016cnb,Boyle:2018tzc,Boyle:2022lyw}.
This avenue is promising, although it relies both on speculative physics near the bang (or bounce), and requires some deft handling of infinity.

Another possibility is to imagine a cyclic cosmology\cite{tolman1931theoretical,steinhardt2002cosmic,lehners2008ekpyrotic,bojowald2008loop,penrose2012basic}.
Typically these involve two novel ingredients: replacing the Big Bang beginning of the conventional picture with a ``bounce'' connecting a crunch and a bang, and a turnover point later in the universe's history where it goes from expanding to contracting. 
This pattern can be repeated indefinitely in both the past and the future.
In the literature, however, most such models are not strictly periodic, with individually symmetric cycles repeating eternally; there is a consistent arrow of time from the infinite (or finite) past to the infinite future, and the crunches and bangs are not time-reverses of each other.

In this paper we present a simple way to get truly periodic cosmological behavior, including very low entropy in the crunch/bounce/bang phase, in the context of a directly quantum framework, where the state of the universe is described by an abstract vector in Hilbert space, evolving unitarily according to the Schr\"odinger equation.
This approach has been called ``state-vector realism," ``Hilbert-space fundamentalism," or ``Mad-dog Everettianism" \cite{Carroll:2018rhc,Carroll2021}.
In this picture, the Hamiltonian is specified simply by its spectrum (the set of energy eigenvalues), rather than through a set of preferred operators.
Emergent semiclassical structures, including space itself, are identified via ``quantum mereology" -- looking for factorizations of Hilbert space in which the spectrum of the Hamiltonian implies the sought-after properties \cite{Albrecht:2010dk,Cotler:2017abq,Carroll:2020gme,Friedrich:2024wps,zanardi2024operational,Adil:2024wok,Soulas:2025kdn,loizeau2025quantum}.
Rather than starting with a classical precursor theory and quantizing, the classical world is recovered as a limit in appropriate circumstances using the standard tools of decoherence and pointer states \cite{zurek2025decoherence}.
This program is still quite preliminary, but some progress has been made in identifying spacetime curvature and gravity \cite{giddings2015,donnelly+giddings2016_2,Cao:2016mst,Bao:2017iye,Cao:2017hrv,Cao:2026uoq}, and some aspects of de~Sitter space have been analyzed  \cite{Banks:2006rx,Beny:2011vh,Czech:2015kbp,SinaiKunkolienkar:2016lgg,Bao:2017qmt,Niermann:2021wco,Cao:2023gkw,Basumatary:2026pyg}.

Cosmological evolution in such an approach depends crucially on whether Hilbert space is infinite- or finite-dimensional.
In this paper we consider the finite-dimensional case.
Although many familiar quantum systems feature infinite-dimensional Hilbert spaces, considerations of Bekenstein/holographic bounds \cite{bekenstein1981,Susskind:1993if,bousso1999,Bao:2017rnv} combined with the de~Sitter entropy \cite{gibbons1977cosmological} provide reasons to believe that a horizon-sized patch of de~Sitter space is described by a finite-dimensional Hilbert space, and (more radically) such a theory could conceivably describe the universe as a whole \cite{Fischler2000,Banks2000,Banks:2000fe,bousso2000,Witten:2001kn,banks+fischler2001,Dyson2002,goheer_etal2003,Parikh:2004wh,Albrecht:2009vr,Banks:2023uit}.
In this picture, we would expect the dimensionality of Hilbert space to be $D\sim e^{S_\mathrm{dS}}$, where the de~Sitter entropy for observed parameters is $S_\mathrm{dS}\sim 10^{122}$.
In this work we are putting aside any inspiration from holography and quantum gravity and simply considering the evolution of a state vector in a finite-dimensional Hilbert space for its own sake.

Unitary evolution in a finite-dimensional Hilbert space is subject to the quantum recurrence theorem: given enough time, any state will return arbitrarily closely to its initial condition \cite{bocchieri1957quantum,schulman1978note,Dyson2002,wallace2015recurrence,chronis2025conditional}.
In an eventual spacetime interpretation, this implies a kind of cyclic cosmology.
However, there is an obvious obstacle to considering such a simple picture as phenomenologically viable: the Boltzmann Brain problem \cite{Dyson2002,Albrecht_2004,Page:2006ys,DeSimone:2008if,Carroll2017}.
If we define an appropriate version of coarse-grained (Boltzmann) entropy, it is possible to choose an ``initial" state of low entropy, and the system will with high probability increase in entropy toward both the past and future from that point; it is natural to identify that moment as the bounce separating a crunch and bang.
But generically, the state will spend the vast majority of its evolution in high-entropy near-equilibrium states.
There will be fluctuations to lower-entropy states, but these are expected to have probabilities $P(\Delta S) \propto \exp(-\Delta S)$, so small fluctuations are enormously more likely than large ones.
As a result, the vast majority of intelligent observers (however those may be defined) in such a cosmology are likely to be minimal fluctuations away from equilibrium, rather than ``ordinary observers'' who arise in the course of thermodynamically sensible evolution from an extremely low-entropy initial condition \cite{Carroll:2008yd}.
Suggested resolutions of this problem have focused on the details of quantum measurement: arguing that physical measurements can never verify the existence of the problematic fluctuations \cite{Banks:2002wr}, or proposing that a suitable coarse-graining can decrease the relative probability of Boltzmann Brains \cite{Albrecht:2009vr}.
(Or, of course, returning to infinite-dimensional Hilbert space, where there are no recurrences \cite{Boddy_2016}.)

In this paper we pursue a different strategy.
We identify a loophole in the argument that small fluctuations are much more common than large ones, thereby restoring the viability of precisely periodic quantum cosmology in a finite-dimensional Hilbert space.

The trick is to carefully choose the energy eigenvalues defining the Hamiltonian of the theory.
In an earlier paper \cite{Carroll:2023mzl} it was noted that if the energy differences $E_i-E_j$ are all commensurable (their ratios are rational, or equivalently they are all integer multiples of some small energy scale), then the evolution is not merely recurrent, but exactly periodic with a finite period that can be much less than the expected recurrence time.
Furthermore, the trajectory is not ergodic in the torus of phases that describes Schr\"odinger evolution; it only traverses a closed one-dimensional loop in that high-dimensional manifold.

As a result, we show here that a carefully chosen low-entropy initial condition, representing an enormous deviation from equilibrium, can periodically repeat on a much shorter timescale than one would expect from $\exp(-\Delta S)$ statistics.
The corresponding cosmology will have relatively fewer small fluctuations, and thus a typical observer will appear in the aftermath of the low-entropy Big Bang, rather than as a random fluctuation in an equilibrium background.
Such cosmologies are therefore phenomenologically viable, and we discuss some of their implications.
(The requirement of commensurable eigenvalues is equivalent to saying that the spectrum is a subset of the spectrum of a simple harmonic oscillator.
In this picture, the theory of everything is essentially a big harmonic oscillator with a very specific initial condition.)

In many ways our discussion harks back to the discourse from the 1890s concerning the recurrence objection to kinetic theory and a statistical understanding of the Second Law.
Poincar\'e proved the (classical) recurrence theorem, showing that systems with bounded phase space eventually return arbitrarily close to their starting point \cite{poincare1890three}. 
He \cite{poincare2003mechanism} and Zermelo \cite{zermelo1896satz,zermelo1896ueber} noted that this raised a problem for any approach that attempted to derive the time-asymmetric Second Law of Thermodynamics from underlying time-symmetric mechanical laws.
Boltzmann responded with a number of scenarios that could escape this dilemma, including the idea of an eternal universe with random fluctuations \cite{boltzmann1895certain,boltzmann1897hrn}.
It was Eddington who pointed out the fatal flaw in this idea, that most fluctuations will be minimally separated from equilibrium \cite{eddington1931end}.
Our cosmological scenario can be thought of as a realization of Boltzmann's aspiration: an eternal universe, with fluctuations, but one for which a particular tremendous fluctuation is responsible for the creation of the large majority of intelligent observers.
We might be finding ourselves in just such a fluctuation, one that has already happened an infinite number of times in the past, and will exactly repeat an infinite number of times in the future.

\section{Setup: Unitary Quantum Mechanics in Finite-Dimensional Hilbert Space}

We start with the idea that Nature is fundamentally quantum-mechanical, and that the appropriate quantum theory of the world can be defined in its own right, rather than resulting from quantization of a classical precursor theory.
We consider a purely unitary (Everettian) picture of quantum mechanics \cite{everett1957relative,wallace2012emergent}, in which the theory is entirely specified by a Hilbert space $\hs$ of finite dimension $\Dim\hs = D$, an initial state vector $|\psi(0)\rangle$, and a Hamiltonian $\widehat{H}$.
Aspects such as observers, measurements, probabilities, and all of the familiar classical structures are to be derived from this simple list of ingredients \cite{Zurek:1994zq,hartle_quasiclassical_2011,Carroll2021}.

The state $|\psi(t)\rangle$ evolves according to the Schr\"odinger equation (setting the reduced Planck constant $\hbar=1$):
\be
\widehat{H}\left|\psi\right\rangle=i\frac{\partial}{\partial t}\left|\psi\right\rangle.
\label{schr}
\ee
For simplicity we will discuss pure quantum states and fundamental time evolution, but it is straightforward to generalize to mixed states or emergent time with an effective Hamiltonian.
Since we are not assuming any particular classical precursor theory, there is no set of preferred observables from which to construct the Hamiltonian.
(In finite-dimensional Hilbert space, every Hermitian operator on the space as a whole is an ``observable"; what gets measured by actual observers depends on how those observers are represented as factors in a tensor-product decomposition of $\hs$, and how the Hamiltonian is expressed in that factorization.)
The only uniquely-defined basis in Hilbert space is the energy eigenbasis, consisting of states $\{|k\rangle\}$ that obey
\be
  \widehat{H}\left|k\right\rangle=E_k\left|k\right\rangle.
  \label{eigenbasis}
\ee
In this basis, the operator $\ham$ is a diagonal matrix, with entries given by the eigenvalues, or spectrum, $\{E_k\}$.
The choice of spectrum entirely defines the quantum theory.

It is straightforward to solve (\ref{schr}) exactly in the energy eigenbasis; the discussion here follows \cite{Carroll:2023mzl}.\footnote{For more details on the geometric picture of Schr\"odinger evolution as happening in a torus of phases, see \cite{kibble1979geometrization,ashtekar1999geometrical,brody2001geometric,bengtsson2017geometry}.}
The solution can be written as a superposition of energy eigenstates,
\be
\left|\psi\left(t\right)\right\rangle=\sum_{k=0}^{D-1}{\alpha_ke^{-i
E_k t}\left|k\right\rangle},
\label{psi-evol}                         
\ee
where the $\{\alpha_k\}$ are constant parameters that can be chosen to be real (by absorbing phases into the definition of $|k\rangle$), with $k\in \{0, 1, 2, \ldots D-1\}$.
We are allowed to shift the energy eigenvalues, $\left\{E_k\right\} \rightarrow\left\{E_k+c\right\}$, since that introduces an irrelevant overall phase. 
Let us order the eigenvalues from lowest to highest, so that $E_k\leq E_{k+1}$. Then we can define shifted energies by extracting the lowest eigenvalue,
\be
\widetilde{E}_k=E_k-E_0,
\ee
and write the state as
\be
|\psi(t)\rangle=\alpha_0|0\rangle+\sum_{k=1}^{D-1} \alpha_k e^{-i \E_kt}|k\rangle.
\ee
The amplitudes stay constant while the phases move linearly in time, and are each periodic, with period $2\pi/\E_k$.
Schr\"odinger evolution can be thought of as straight-line motion on a $(D-1)$-dimensional torus, with coordinates given by $\tilde\theta_k(t) = \E_k t$.

In the generic case where the ratio of any two energies $\E_k/\E_j$ is irrational, and all of the basis states have nonzero amplitudes, the trajectory will never exactly come back to its origin, but it will be dense in the $(D-1)$-dimensional torus, consistent with the recurrence theorem.
In the special case where the ratio of any two energies is a rational number, $\forall (k,j): \E_k/\E_j \in \mathbb{Q}$, the set of energies is commensurable, and we can write
\be
  \E_k = p_k \omega_*,
  \label{pk}
\ee
where $\omega_*$ is some fiducial energy scale and $\{p_k\}$ is a set of coprime integers.
Then after a time
\be
  T_* = \frac{2\pi}{\omega_*},
  \label{Tstar}
\ee
each phase will increment by $\E_k t \rightarrow \E_k t + 2\pi p_k$, and the state vector will return to precisely where it started, describing periodic evolution with period $T_*$ \cite{bocchieri1957quantum}.

There are two important things to note about the condition that shifted energies be commensurable.
First, given that all energies are multiples of some fiducial value as in (\ref{pk}), our spectrum is just that of (some subset of) a simple harmonic oscillator.
That makes sense, as it is the requirement that different energies evolve with commensurate frequencies.
This state of affairs is less restrictive than we might originally think, since we only need some subset of the harmonic oscillator spectrum, not necessarily the whole thing.
Given any finite set of real numbers $\{x_k\}$, there exists a real number $\omega_*$ and a set of integers $\{p_k\}$ that approximates it arbitrarily well; i.e. $\forall \epsilon > 0, \exists \omega_*, \{p_k\} : \sum_k |x_k - p_k\omega_*| < \epsilon$.
If we want our spectrum to ultimately support a factorization of Hilbert space into local subsystems and other requirements, we can come as close to the required conditions as we like.

The other thing to note is that this periodicity timescale is significantly shorter than a standard recurrence time.
We see that $T_*$ is set by the fiducial energy $\omega_*$, with no explicit dependence on the size of the overall Hilbert space.
By contrast, the recurrence time to enter a macrostate represented by a Hilbert subspace of dimension $d$ is expected to be $T_\mathrm{recur}\sim e^{c(D-d)}/\delta E$, where $\delta E$ is a typical level spacing (perhaps of order $\omega_*$) and $c$ is a constant that depends on details of the spectrum and the criteria for recurrence.
(See \cite{chronis2025conditional,Gupta:2026tkz} for a more detailed discussion.)
This is clearly much longer than $T_*$, especially considering $D \sim \exp(10^{122})$.
This will be discussed in more detail in Section~\ref{entropy_fluctuations}.

Of course, a compelling cosmological model should specify not only the evolution of the state vector, but also a way to interpret that quantum state in terms of spacetime and quantum fields.
We cannot offer that, although some tentative steps have been taken in that direction, as mentioned in the Introduction.
But if the universe is described by unitary quantum mechanics in a finite-dimensional Hilbert space, the set of possible evolutions is highly constrained, and from the evolution of the state vector itself we can infer likely behaviors for the classical spacetime interpretation.
We discuss this connection in Section~\ref{spacetime_interp}.

\section{Quantum Boltzmann Entropy}\label{quant_therm_ent}

In asking whether such a setup can serve as the basis for a phenomenologically realistic description of our universe, recovering an appropriate spacetime interpretation is not the only relevant consideration. Another consideration of prime importance is whether we can recover an appropriate thermodynamic interpretation, given that our observed universe started out in an extremely low-entropy state and has been increasing in entropy ever since, as per the Second Law. For this, we fortunately do not need a fully worked out solution to the problem of emergence. We can learn a lot about thermodynamic behaviors squarely within the Hilbert-space framework, in which a statistical treatment of quantum states allows us to investigate entropic evolution. 

First, we have to define what we mean by ``entropy'' in the context of quantum thermodynamics.
Constructing the density operator corresponding to our state,
\be
  \hat\rho = |\psi\rangle\langle\psi|,
\ee
a reasonable initial attempt is to calculate the von~Neumann entropy, 
\be
  S_\mathrm{vN} = -\trace(\hat\rho\log\hat\rho).
\ee
\noindent But since the universe is evolving unitarily (i.e., it is a closed system), $S_\mathrm{vN}$ is conserved. Furthermore, if the state is pure, then $S_\mathrm{vN}$ vanishes. It is thereby not the right quantity that is relevant for thermodynamics, at least when closed systems are involved. 

Instead, let's define a ``quantum Boltzmann entropy,'' in the sense of the classical $S_\mathrm{B}=k_B \log W$ formula that appears on Boltzmann's tombstone. The idea classically is to partition phase space into macrostates, regions $\Omega_a$ with volume $W_a$, and then to assign each phase-space point (i.e., microstate) an entropy based on the volume of the macrostate that it is in. The equilibrium macrostate, which corresponds to the region of largest volume, should have maximum entropy. The classical Boltzmann entropy is thus defined as proportional to the logarithm of macrostate volume, $S_\mathrm{B}=k_B\log W_a$.

Advantages to the Boltzmannian approach are that once a coarse-graining is decided, $S_\mathrm{B}$ is always positive, and it changes with time, even for closed systems. As the system's phase space trajectory traverses different regions, its entropy at any given time is determined by the associated macrostate. For the vast majority of trajectories starting out in a low-entropy macrostate, $S_\mathrm{B}$ tends to increase toward the maximum-entropy macrostate and overwhelmingly stay there, exactly as a statistical Second Law would predict.  

These properties contrast with the classical Gibbs-Shannon entropy, $S_\mathrm{G} = - \sum_i p_i \log p_i$, where $p_i$ indicates the probability that the system is in microstate $i$. As we saw with its quantum counterpart, von~Neumann entropy, $S_\mathrm{G}$ is conserved for closed systems, and it vanishes for an exact microstate with probability 1, no matter which microstate that is. Thus, the same objection holds for $S_\mathrm{G}$ as it does for $S_\mathrm{vN}$; it is not the right quantity for the thermodynamics of closed systems. 

In many ways, the Gibbs and von~Neumann entropies are easier to manipulate mathematically, and they adapt well to the study of open systems, which is why they are widely used in statistical mechanics. Their problems for the Second Law can be circumvented through an alternative coarse-graining procedure of repeatedly updating the probability distribution by throwing away information \cite{ehrenfest1990conceptual}. However, this strategy works precisely because it relies on an implicit coarse-graining into macrostates. 
Whenever we throw away information, we do so on the basis of the system's macrostate at that time. 

Ultimately, the Boltzmann entropy is conceptually prior and underpins the Second Law for closed systems. Most importantly, it's the Boltzmann entropy that actually enters into explanations of the arrow of time, as knowing that the universe's Boltzmann entropy was low in the past (the Past Hypothesis) serves to anchor present-day inferences about memories and causality, not to mention aging and the psychological arrow of time \cite{albert2003time,carroll2010eternity,Fernandes_2023}.

We will therefore define a quantum analogue of the classical Boltzmann entropy, modifying earlier proposals \cite{hemmo2012road,goldstein_gibbs_2020,safranek_brief_2021}. The role of coarse-graining is played by partitioning Hilbert space into a direct sum of macrostate subspaces:
\be
  \hs = \bigoplus_a \hs_a.
\ee
The equivalent of the phase-space volume of a classical macrostate is the dimensionality of the subspace $\hs_a$, namely $d_a = \dim \hs_a$.

If our pure state were strictly inside one particular subspace $\hs_{\bar a}$, it would be natural to assign a quantum Boltzmann entropy of $S_{\bar a}=\log d_{\bar a}$. But a typical state will have nonzero components in many such subspaces. Let $\hat \Pi_a$ be a projection operator onto $\hs_a$. Then the probability that, were we to do a projective measurement onto the various macrostates, we would observe $\hs_a$ for a state $|\psi\rangle$ is
\be
  p_a = \langle\psi | \Pi_a |\psi\rangle.
\ee
So the natural way to define the global quantum Boltzmann entropy is
\be
  S_\mathrm{QB} = \sum_a p_a \log d_a,
\ee
setting the Boltzmann constant $k_B = 1$. Since $\log d_a$ is simply the quantum Boltzmann entropy $S_a$ of the macrostate associated with subspace $\hs_a$, $S_\mathrm{QB}$ is straightforwardly the weighted average of component $S_a$'s. $S_\mathrm{QB}$ thus captures the novelty of macrostate superpositions.

Given that we never encounter macroscopic superpositions, however, one may wonder whether $S_\mathrm{QB}$ is relevant phenomenologically, i.e., to recover a classical-looking thermodynamic arrow. It turns out that it is if we want the Past Hypothesis to have predictive power for a quantum cosmology. Classicality emerges when the global state decoheres into non-interfering branches, but the thermodynamic arrow is the consequence of a low-entropy initial state. Therefore, in order to satisfy both constraints, the ``Quantum Past Hypothesis'' must stipulate that the initial global state decoheres into low-entropy branches. This condition is definitively met by minimizing $S_\mathrm{QB}$, which ensures that all branches evolve from the lowest-entropy macrostate.  
 
The Quantum Past Hypothesis thereby aligns entropic behavior at two levels -- the global level and the branch level -- both of which exhibit a Second law. At the global level, the Second Law means that $S_\mathrm{QB}$ tends to increase. This is a distinctly quantum result with no classical analogue, since the global state is usually in a superposition of different macrostates. Decoherence then transforms superpositions into branching structure. The classical Second Law at this emergent level means that the entropy of a branch tends to increase. 

Let us define a ``branch thread" as a particular semiclassical history of the universe.
At any one moment each branch thread is associated with a particular branch, but at decoherence events where different macroscopic outcomes are realized, different threads will head down different future branches.
The proliferation of branches yields many thermodynamic arrows, each of which is associated with a unique branch thread. 
Though it's only along a particular branch thread that classically emergent systems, including observers like us, ever experience a thermodynamic arrow, the Quantum Past Hypothesis ensures that branch threads typically have a thermodynamic arrow.  

To see how this works, let's compare the entropic behavior of decohered branches (both in terms of quantum and classical Boltzmann entropy) with that of the global state. Decoherence occurs under an appropriate decomposition into subsystems, where one is the target system and the other is the environment whose states are mutually orthogonal. Every Hilbert macrostate thus factorizes into a tensor product: $\hs_a = \hs_S^{(a)} \otimes \hs_E^{(a)}$, and the Hilbert space decomposition is
\be
 \hs = \bigoplus_a \left[\hs_S^{(a)}\otimes \hs_E^{(a)}\right].
\ee 

The need for different environment factors on different branches might be important when branches have very different spacetime structures, for example where a false vacuum has decayed on one branch but not the other.
But for now let's focus on a more everyday regime, where for any classical system of interest (e.g., galaxies, stars, planets, etc.), we can trace out the \emph{same} environment. 
In that case we can write the Hilbert space decomposition as
\be
 \hs = \left[\bigoplus_a \hs_S^{(a)}\right]\otimes \hs_E.
\ee
Here, the environment factor $\hs_E$ is taken to be common among all of the possible system states, while $\hs_S^{(a)}$ is the $a$th macrostate of the system degrees of freedom. So we have $\hs_a = \hs_S^{(a)}\otimes \hs_E$ and $d_a = d_S^{(a)}d_E$. 
We are not keeping track of the environment's macrostate other than the fact that it records system states upon interaction. 

Let $\{|\phi_{n_a}^{(a)}\rangle\}$ be the pointer basis for $\hs_S^{(a)}$, where $n_a$ indexes the basis states of $\hs_S^{(a)}$, and runs from $1$ to $d_S^{(a)}$. As interactions entangle the system and environment, a decomposition into branches looks like
\be
 |\psi\rangle = \sum_{a, n_a} \psi_{a, n_a} |\phi_{n_a}^{(a)}\rangle\otimes |e_{a, n_a}\rangle.
\ee
Classical-looking systems emerge because interference between branches is eliminated by the mutually orthogonal environment states, and they continue to evolve autonomously (until they recohere in the future, after the thermodynamic arrow has given out).

Each decohered branch is thus in a definite thermodynamic macrostate, and the branch-relative quantum Boltzmann entropy of a branch is $S_a = \log d_S^{(a)} + \log d_E$. 
When $S_a$ is less than the maximum, it increases along a branch thread due to the overwhelming likelihood of the same branch or descendant branches ending up in larger subspaces corresponding to higher entropy macrostates. However, it's important to qualify that whereas the $\log d_S^{(a)}$ term carries macrostate information and can change over time, the $\log d_E$ term stays constant by construction, so the thermodynamic contribution of $S_a$ is really only coming from $\log d_S^{(a)}$. 

This is the Second Law operating at the branch level, and it should mirror the classical Second Law. In other words, an increase in the quantum Boltzmann entropy $S_a$ should also manifest as an increase in the classical Boltzmann entropy $S_\mathrm{B}$. Here's the big step to relate the two. Let the pointer state represent $N$ particles. By the semiclassical correspondence between phase space volume and Hilbert space dimensionality \cite{Landau:1980mil}, each pointer basis state $|\phi_{n_a}^{(a)}\rangle$ can be thought of as occupying a classical phase space volume of
\be
  h^{3N},
\ee
where $h$ is Planck's constant. Then the $d_a$ basis elements spanning $\hs_a$ will tile a phase-space volume
\be
  W_a = d_a h^{3N} = d_S^{(a)}d_E h^{3N}.
\ee
This is just the quantity that appears in the formula for the classical Boltzmann entropy, $S_\mathrm{B} = k_B \log W_a$, which increases whenever $d_S^{(a)}$ increases, as expected.

Nevertheless, it's not enough to recover the thermodynamic arrow within individual branches or branch threads. The thermodynamic arrow should be typical across branches. For this we need a measure, which is none other than the Born measure embedded in the global quantum state. The evolving Born probabilities $p_a$ that enter into the global quantum Boltzmann entropy, $S_\mathrm{QB} = \sum_a p_a \log d_a$, thus reflect how subsets of branches evolve. The Quantum Past Hypothesis makes use of the Born measure precisely to guarantee that the thermodynamic arrow is typical across branches. 

The Quantum Past Hypothesis establishes that the global state at some initial time is concentrated wholly in the lowest entropy macrostate; $p_a = 1$ for minimal $d_a$ entails that all branches start out in the smallest subspace. 
(This strict condition could be relaxed, but we assume it here for simplicity.)
Then $S_\mathrm{QB}$ subsequently increases as the probability distribution evolves, provided of course that we are not talking about energy macrostates; otherwise, the probability distribution would be time independent. Note also that the subspace dimensionalities $d_a$ are fixed by the original Hilbert space partitioning, so $S_\mathrm{QB}$ increases only as the probability distribution spreads out over different macrostates. During equilibration, the probability distribution progressively favors higher entropy macrostates, leading to more branches ending up in those larger subspaces.  

However, because the global state remains in a superposition under unitary evolution, it also will sample lower entropy macrostates. The larger the finite-dimensional Hilbert space, the closer the probability approaches one of ending up in the highest entropy macrostate. Nevertheless, the global state's $S_\mathrm{QB}$ will always be less than the maximum $S_a$. So by the time equilibrium is reached, the value of $S_\mathrm{QB}$ reveals that a tiny fraction of branches under the Born measure never reach the highest entropy state, though the vast majority of branches do. 

One might now worry that $S_\mathrm{QB}$ increases predominantly because the global state decoheres into an increasing number of branches, rather than because entropy increases within the branches (or branch threads), but that's not so. Consider a small quantum system that decoheres. Initially in a pointer state $|\phi_0\rangle_S$ at $t_0$, it evolves a short time later at $t_1$ into a superposition $\sum_i\beta_i|\phi_i\rangle_S $, where $i$ counts the basis states of the entire Hilbert space $\hs$ and runs from $1$ to $\Dim\hs$. The evolution with branching appears schematically as
\be
  |\phi_0\rangle_S |e_0\rangle_E
  \rightarrow \sum_i \beta_i |\phi_i\rangle_S|e_i\rangle_E.
\ee

The quantum Boltzmann entropy will therefore evolve as
\be
  S_\mathrm{QB}(0) = \log d_0 \rightarrow S_\mathrm{QB}(1) = \sum_i |\beta_i|^2 \log d_{a(i)}.
\ee
Here, $d_{a(i)}$ is the Hilbert space dimension of the macrostate $\hs_a$ of which the pointer state $|\phi_i\rangle_S$ is a member. But since the branching represents the decoherence of a presumably small quantum system in superposition, and because $t_1 - t_0 = \delta t$ is presumably insufficient time for the system to have evolved into a radically new macrostate in any one of its branches, we expect to have $d_0 \approx d_{a(i)}\ (\forall i)$, up to tiny quantum fluctuations. 
Furthermore, given that $\sum_i |\beta_i |^2= 1$, to a good approximation we have
\be
  S_\mathrm{QB}(0) \approx S_\mathrm{QB}(1).
\ee

This result is independent of the number of branches into which the state decohered, which reassures us that increasing $S_\mathrm{QB}$ comes from increasing coarse-grained entropy $S_a$ within each branch (or branch thread), not an increasing number of branches. This result also highlights how the thermodynamic arrow is distinct from other temporal asymmetries involving entanglement, decoherence, and measurement uncertainty, in contrast to existing proposals of quantum thermodynamic entropy, which end up conflating these phenomena. 

For one, it has been proposed that the ``thermal entropy'' \cite{Susskind:2005js} of a closed quantum system is derived by coarse-graining it into interacting subsystems and adding up their entanglement von Neumann entropies: $S_{th} = -\sum_i \trace(\hat\rho_i\log\hat\rho_i)$. However, $S_{th}$ cannot be the entropy figuring into the thermodynamic arrow because it increases merely with growing entanglement and the proliferation of branches, like during the interval $\delta t$ that a small quantum system decoheres. Moreover, since $S_{th}$ simulates a closed quantum system as a collection of open systems, it increases monotonically and cannot fluctuate out of equilibrium, though the thermodynamic entropy of the universe does fluctuate under a statistical Second Law. In contrast, $S_\mathrm{QB} = \sum_a p_a S_a$ disambiguates the appropriate coarse-graining procedure for closed quantum systems (recall that $S_a = \log d_a$), so the Second Law emerges in all the right ways.

Another proposal called ``observational entropy'' \cite{safranek_etal:2019, safranek_brief_2021} has been presented as a Boltzmannian definition of quantum thermodynamic entropy: $S_O = -\sum_a p_a \log p_a + \sum_a p_a S_a$. It is actually a hybrid quantity containing not only the Boltzmannian $S_\mathrm{QB}$ term but also a Gibbsian term, $ -\sum_a p_a \log p_a$, which is supposed to capture the uncertainty of measuring any given macrostate. 
This term quantifies the possibility that an observer doesn't know what macrostate they are in; it does not apply to our cases of interest, the evolution of the global state and the thermodynamic arrow within branch threads.

A similar alternative to observational entropy was presented in \cite{goldstein_gibbs_2020}. They proposed to treat quantum Boltzmann entropy not as a scalar but as an operator: $\hat{S}_\mathrm{QB} = \sum_a \Pi_a S_a$. On the face of it, $\hat{S}_\mathrm{QB}$ seems to be compatible with $S_\mathrm{QB}$ considering that $S_\mathrm{QB} = \sum_a p_a S_a$ is just the expectation value. However, \cite{goldstein_gibbs_2020} reject this move, largely because of their skepticism about the existence of long-lived macrostate superpositions.
From the Everettian perspective taken in this paper, such superpositions are inevitable, and taking the expectation value to obtain a scalar entropy produces the quantity relevant to us.


What these competing Boltzmannian definitions have in common is that they are primarily concerned with the probability of ending up in definite macrostates or individual decohered branches, as opposed to with properties of the global quantum state. These proposals are thus not aimed at describing the thermodynamic arrow of the whole universe, whereas our simple formula is able to do that at both the branch and global levels. The Quantum Past Hypothesis ultimately requires minimizing $S_\mathrm{QB}$ in order to recover the Second Law as a typical occurrence. 
 
\section{Entropy Fluctuations and Boltzmann Brains}\label{entropy_fluctuations}

With this definition of quantum Boltzmann entropy in hand, we return to considering evolution in a Hilbert space of fixed dimensionality, where the energy eigenvalues are rational multiples of each other (i.e., integer multiples of some smallest energy factor), so that the unitary evolution is periodic with some known timescale.
In this setup, we might expect the quantum Boltzmann entropy to spend almost all its time near its maximum value, with downward fluctuations occurring with probability $P(\Delta S) \propto \exp(-\Delta S)$, as we have in the classical case.
If that were the case, even if the state occasionally fluctuated to very low entropy, a typical downward fluctuation in entropy that satisfied some given condition (e.g. ``containing an intelligent observer'') would be a minimal such fluctuation.
This would lead to the Boltzmann Brain problem, as foreseen in this precise context in \cite{Carroll:2008yd,Carroll:2023mzl}.

But it is not the case once we specialize to commensurable energy eigenvalues, because the evolution (\ref{psi-evol}) is not ergodic in any sense.
It travels through a torus subspace within Hilbert space, and the coefficients $\alpha_k$ remain fixed.
With incommensurable eigenvalues, the evolution is ergodic within that torus, eventually coming arbitrarily close to any specified point on it.
But with commensurable eigenvalues, the evolution describes a closed one-dimensional curve, and is highly non-ergodic.
The long-term trajectory does not sample the whole torus fairly.
As a result, if we choose to start in an (admittedly non-generic) low-entropy initial condition, we can return exactly to that initial condition well before we sample a large number of Boltzmann-Brain-like fluctuations.

\begin{figure*}[t]
\begin{center}
\includegraphics[width=1\textwidth]{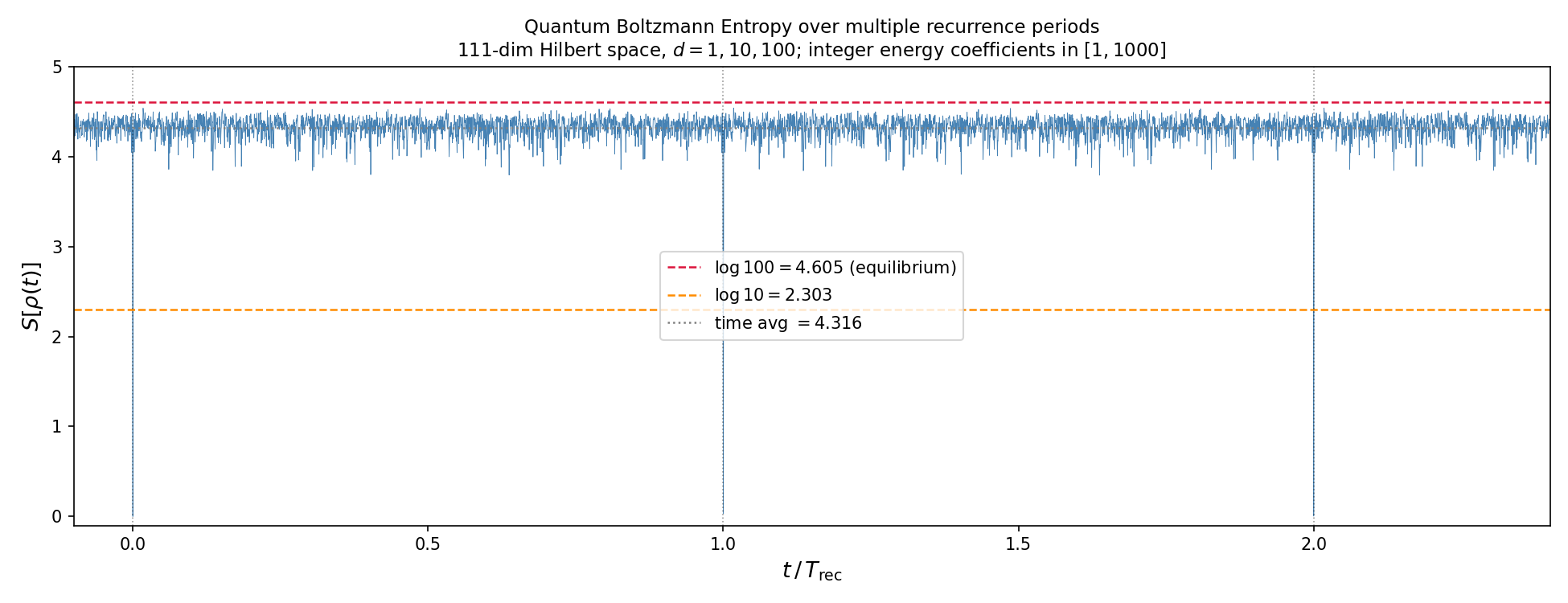}
\caption{Evolution of quantum Boltzmann entropy in a finite-dimensional Hilbert space with exact recurrence and a low-entropy condition at $t=0$. 
These plots use $D=111$, with three macrostates of dimensions 1, 10, and 100, and energy eigenvalues given by random integers between 1 and 1000 times a fiducial energy $\omega_*$.
The vertical-looking lines at $t/T_\mathrm{rec} = 0, 1, 2...$ are downward excursions to the minimum-entropy macrostate.}
\label{entropy_evolution}
\end{center}
\end{figure*}

To illustrate this loophole, we have simulated a situation with a 111-dimensional Hilbert space partitioned into three macrostates, $\hs = \hs_1 \oplus \hs_{2} \oplus \hs_3$, with dimensions $d_1 = 1$, $d_2 = 10$, and $d_3=100$.
The macro-spaces are chosen to be Haar-random with respect to the energy eigenstates.
We chose all of the energy eigenvalues to be integer multiples of some fixed energy $\omega_*$, so that $\widetilde{E}_k = n_k\omega_*$, with the $n_k$s chosen as random integers between 1 and 1000.
Then we started with an ``initial'' state entirely within the smallest macro-space, $|\psi(0)\rangle \in \hs_1$, corresponding to $S_\mathrm{QB}=0$, and let it unitarily evolve.

Several periods of the resulting evolution of the quantum Boltzmann entropy are shown in Figure~\ref{entropy_evolution}.
As expected, most of the time is spent in an equilibrium macrocondition with nearly-maximum entropy, plus small fluctuations downward.
The achieved highest entropy is slightly smaller than the nominal maximum entropy, since the state includes contributions from lower-entropy macrostates; as the dimensionality of Hilbert space is increased, this discrepancy would be negligibly small.
Furthermore, the initial condition represents a huge and atypical downward fluctuation in entropy.
The evolution of this periodic system, therefore, includes one enormous entropy fluctuation per cycle, and a relatively small number of smaller fluctuations; the repetition frequency of the large fluctuations is much greater than we would expect on the basis of $P(\Delta S) \propto \exp(-\Delta S)$.

\begin{figure*}[t]
\begin{center}
\includegraphics[width=1\textwidth]{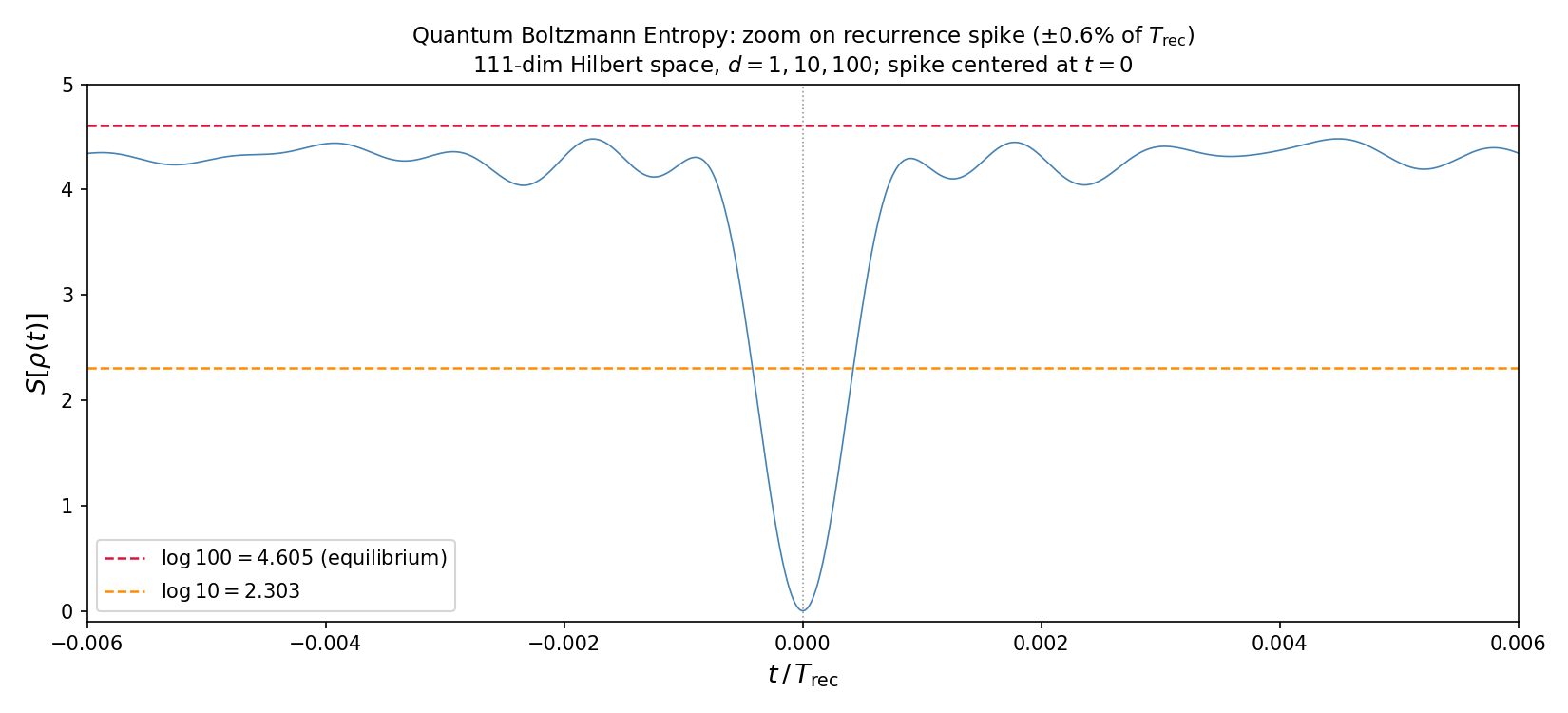}
\caption{Same simulation as Figure~\ref{entropy_evolution}, but now we have zoomed in near the low-entropy excursion around $t=0$.
The evolution of $S_\mathrm{QB}(t)$ is approximately symmetric around the minimum, but not exactly so.
In our interpretation, the minimum represents the bounce, with arrows of time directed away from it in either direction.}
\label{entropy_zoom}
\end{center}
\end{figure*}

Figure~\ref{entropy_zoom} zooms in on the neighborhood of the low-entropy initial condition.
Moving from left to right, the curve represents a crunch/bounce/bang process; but because entropy is minimized at the bounce, observers on both sides would interpret it locally as a ``bang," with the arrow of time pointing away from the entropy minimum.

Note that the entropy is not necessarily precisely symmetric around the minimum.
To see this, remember that $S_\mathrm{QB} = \sum_a p_a\log d_a$, where $d_a$ is the dimensionality of the macro-space $\hs_a$, and in a pure state $|\psi\rangle$ we have $p_a = \langle \psi |\hat\Pi_a |\psi\rangle$, where $\hat\Pi_a$ is the projector onto $\hs_a$.
In the energy eigenbasis we have $|\psi(t)\rangle = \sum_k \alpha_k e^{-iE_k t}|k\rangle$.
Define the matrix elements of the projectors in this basis as
\be
  M^{(a)}_{kk'} = \langle k |\hat\Pi_a |k'\rangle.
\ee
Then
\be
  p_a(t) = \sum_{k, k'} \alpha_k^* \alpha_{k'} e^{i\left(E_k-E_{k'}\right) t} M_{k k'}^{(a)}.
  \label{pas}
\ee  
Replacing $t \rightarrow -t$, we have
\be
  p_a(-t)=\sum_{k, k'} \alpha_k^* \alpha_{k'} e^{-i\left(E_k-E_{k'}\right) t} M_{k k'}^{(a)}.
\ee
In order to guarantee that $S_\mathrm{QB}(t) = S_\mathrm{QB}(-t)$, we need to have $p_a(t) = p_a(-t)$, which will only happen if every Fourier coefficient with frequency $\omega$ matches.
That condition is
\be
  \sum_{E_k - E_{k'}=\omega} \alpha^*_k \alpha_{k'} M_{kk'}^{(a)}
  = \sum_{E_k - E_{k'}=-\omega} \alpha^*_k \alpha_{k'} M_{kk'}^{(a)}
  = \sum_{E_k - E_{k'}=\omega} \alpha^*_{k'} \alpha_{k} M_{kk'}^{(a)*}.
\ee
The sums are taken over all $k, k'$ that satisfy the indicated condition.
The first expression comes from $p_a(t)$, the second from $p_a(-t)$, and in the third we have relabeled dummy indices $k \leftrightarrow k'$ and used Hermiticity of the projectors, $M_{k' k}^{(a)}=\left(M_{k k'}^{(a)}\right)^*$.
We are left with a reality condition that must hold if $S_\mathrm{QB}(t) = S_\mathrm{QB}(-t)$:
\be
\mathrm{Im}\left(\sum_{E_k - E_{k'}=\omega}
\alpha_k^* \alpha_{k'} M_{k k'}^{(a)}\right) =0 \quad \text { for all } \omega, \ a.
\ee
This will not be satisfied for an arbitrary choice of a low-entropy initial condition and macrostate projectors, so the entropy curve is not precisely symmetric around the minimum.
(We can choose the $\alpha_k$'s to be real by absorbing phases into the definitions of the energy eigenstates, but that will generally introduce phases into the matrix elements $M^{(a)}_{kk'}$.)

Nevertheless, interpreting the low-entropy point as a crunch/bounce/bang transition, the classical thermodynamic entropy will increase on typical branches as we move away from the low-entropy point toward either the past or the future.
Ordinary observers will be able to come into existence on both sides of the bounce; either set of observers ($t>0$ and $t<0$) will experience conventional thermodynamic evolution, labeling ``the past'' as the direction in which the entropy of the universe is smaller, although they will characterize each other as having reversed arrows of time (a fact which is of course completely unobservable, as we cannot see through the Big Bang).

In addition to these ordinary observers, there can still be Boltzmann Brains (in the generalized sense of ``random dynamical fluctuations satisfying some specified conditions to count as an observer") in the near-equilibrium phase.
Such fluctuated observers will be relatively rare compared to those expected in the standard picture of ergodic evolution and Poincar\'e recurrences, however, just because the time between exactly returning to the low-entropy state is much smaller than the ordinary recurrence time.
Therefore, this scenario has a chance of representing an eternal periodic oscillating universe that avoids the Boltzmann Brain problem.

Let's be a bit more quantitative about this last point, by comparing the number of expected ordinary observers to the number of expected Boltzmann Brains (BBs).
For what range of parameters will the period be short enough to avoid domination by BBs?
Let $N_\mathrm{OO}$ be a typical number of ordinary observers that would be produced over a cosmological history resembling our observable universe.\footnote{There would actually be enormously more than this number, if we were to account for observers on all branches of the wave function.
But such observers should also be weighted by the amplitude-squared of the branches that they are on, which compensates, so a standard expectation value in a single stochastic universe is what we want.}
In \cite{Carroll2017} this was estimated as $10^{11} \leq N_{\mathrm{OO}}\leq 10^{124}$; the precise value will be essentially irrelevant.
With this we compare the number of BBs produced in a de~Sitter Hubble volume $H_*^{-3}$ over a time $T_*$:
\be
  N_\mathrm{BB} = \Gamma_\mathrm{BB} T_* H_*^{-3},
  \label{NBB}
\ee
where $\Gamma_\mathrm{BB}$ is the rate of fluctuations into BBs (however the relevant conception may be defined) per spacetime volume.
This can be estimated as 
\be
  \Gamma_\mathrm{BB} \sim e^{-S_\mathrm{BB}} \sim e^{-M_\mathrm{BB}/T_\mathrm{dS}} \sim e^{-10^{66}},
\ee
where $S_\mathrm{BB}$ is the entropy associated with a BB, $M_\mathrm{BB}$ is the mass of a BB (assumed to contain of order Avogadro's number of baryons) and $T_\mathrm{dS} = 10^{-33}\,$eV is the de~Sitter temperature.
The units of such quantities are, again, essentially irrelevant; indeed, the numerical value of the spatial volume $H_*^{-3}$ in (\ref{NBB}) is likewise irrelevant, in the sense that changing units from Planck times to Hubble times doesn't alter the numerical coefficient.
Keeping only numerically relevant quantities, the criterion that our universe not be dominated by Boltzmann Brains, $N_\mathrm{BB} < N_\mathrm{OO}$, is that the period (\ref{Tstar}) satisfy
\be
  T_* = \frac{2\pi}{\omega_*} \leq e^{10^{66}}.
  \label{Tstarmax}
\ee

To judge whether this is reasonable, we need to estimate, or at least bound, sensible values of the energy scale $\omega_*$.
Here we should admit that we have little to go on, especially as the time parameter that appears in the Schr\"odinger equation (\ref{schr}) and the energies that appear in the spectrum of $\widehat{H}$ might not be directly interpretable in our emergent spacetime description; they could instead be related by some nontrivial (holographic, non-local) map that at this point is not very well understood.
With that caveat in mind, a plausible upper bound for $\omega_*$ comes from imagining that the minimal energy difference will not be larger than the energy of a single photon with a wavelength of order the de~Sitter Hubble radius, or $\omega_* \sim 10^{-33}\,$eV.
That corresponds to a period of not much more than a Hubble time, $T_* \sim 10^{10}\,$yr.
That in turn implies a Big Crunch at a future time that is relatively soon, cosmologically speaking, but this is only a lower bound on the period.

As an upper bound to $T_*$, we might suppose that $\omega_*$ is simply some ``large'' energy scale (the magnitude is, one more time, essentially irrelevant) divided by $D \sim e^{10^{122}}$, the dimensionality of Hilbert space.
That corresponds to a period $T_* \sim e^{10^{122}}$, which is much \emph{longer} than the upper bound from Boltzmann Brain domination in (\ref{Tstarmax}).

The question then becomes, is it reasonable to imagine that the actual period is much smaller than this upper bound, enough to avoid BB domination?
We believe that it is entirely reasonable.
The corresponding energy scale, $\omega_* \sim 1/D$, is fantastically tiny in comparison to any scale from realistic particle physics or cosmology.
Moreover, even if there are $D$ energy eigenstates, there could easily be enormous degeneracies in the actual spectrum, indicating the presence of global symmetries, so a value of $\omega_*$ that is large compared to $1/D$ doesn't imply that the maximum energy is anywhere close to $D$.
We therefore conclude that, while our considerations here are somewhat impressionistic, it should be completely allowed for a model of this form to have a period in the allowed range
\be
  10^{10}\,\mathrm{yr} \leq T_* \leq e^{10^{66}}.
  \label{Tbound}
\ee

We should note the different stance toward fluctuations in a finite-dimensional Hilbert space advocated by \cite{Banks:2002wr}.
They take a Copenhagen view on quantum measurement rather than the Everettian perspective considered here. 
Accordingly they argue that quantum mechanics requires measuring devices that are ideal only in the limit of large apparatus mass, which becomes problematic in the presence of gravity, and therefore that Poincar\'e recurrences are operationally meaningless.
From an Everettian view, by contrast, there will be fluctuations in the global quantum state representing effectively-decoherent entanglement between systems and apparatuses that are macroscopically large but not infinitely big, and there is no reason to discount such fluctuations when taking an inventory of observers in the universe.
It is possible that a careful analysis similar to \cite{Banks:2002wr} could be carried out in a strictly Everettian framework, and it might suggest that true observers require larger fluctuations than the $S_\mathrm{BB}\sim M_\mathrm{BB}/T_\mathrm{dS}$ value that we have studied here, and this could loosen the bound in (\ref{Tbound}), perhaps by an appreciable amount; we do not take up this challenge here.

\section{Spacetime Interpretation}
\label{spacetime_interp}

A truly successful theory of a cyclic universe would have not only an acceptable evolution of the thermodynamic entropy, but also a clear spacetime interpretation.
At this time we do not have a well-defined map from states $|\psi\rangle$ in our finite-dimensional Hilbert space to a semiclassical spacetime (when such a thing exists), although there has been some progress in this direction \cite{Cao:2016mst,Cotler:2017abq,Cao:2017hrv,Bao:2017iye,Bao:2017qmt,Carroll:2020gme,Niermann:2021wco,Cao:2023gkw,zanardi2024operational,loizeau2025quantum,Cao:2026uoq,Basumatary:2026pyg}.
The basic technique is quantum mereology: finding a factorization of Hilbert space $\hs = \bigotimes_n \hs_n$ that has the right properties (locality, localization near classical trajectories, robustness under environmental monitoring) to allow for the emergence of familiar quasi-classical structures.

Even without a complete picture, however, we can reason from fairly general features of the cosmological evolution of our model to what we expect the corresponding spacetime to look like, if such a map is eventually derived.
The salient features of a would-be spacetime interpretation are these:
\begin{enumerate}
    \item The low-entropy minimum represents a bounce connecting a Big Crunch on one side to a Big Bang on the other.
    Call the bounce phase $B$.
    \item On each side of the bounce we have conventional cosmological evolution from the high-temperature/low-entropy phase toward an asymptotically de~Sitter phase.
    Call these phases $\Lambda^\pm$, indicating conventional $\Lambda$CDM cosmology.
    \item Much of the time is spent in a high-entropy equilibrium phase whose closest spacetime interpretation is empty de~Sitter.
    Call this phase $E$ for equilibrium.
\end{enumerate}

\begin{figure}[t]
\begin{center}
\includegraphics[width=0.65\textwidth]{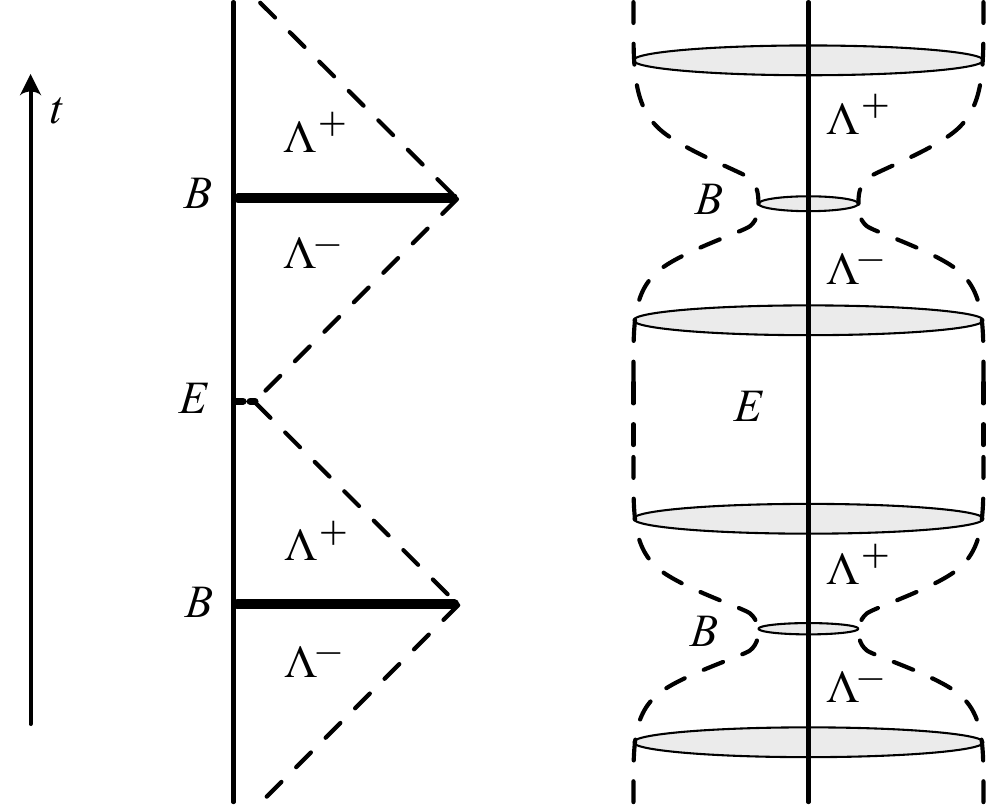}
\caption{A conjectural spacetime diagram for a spacetime interpretation of the quantum cyclic universe, with time running vertically.
The left figure is a conformal diagram, while on the right we have a schematic representation of the Hubble volume surrounding an observer at the origin.
$B$ is the bounce phase, $\Lambda^\pm$ are the expanding/contracting $\Lambda$CDM phases (with respect to a time coordinate running vertically upward), and $E$ is the de~Sitter equilibrium phase.
The diagonal lines at the edges of $\Lambda^\pm$ do not represent a conformal boundary (such as $\mathcal{I}$), but rather the edge of the causal diamond accessible to an observer at the origin, represented by the left vertical line.
Phase $E$ lasts for much longer than is suggested by the pictures.}
\label{st-diagram}
\end{center}
\end{figure}

A possible spacetime diagram implementing these features is shown in Figure~\ref{st-diagram}, first as a conformal diagram and then as a sketch of the horizon volume associated with a chosen geodesic.
At the moment it is conjectural, but should capture the essential properties of this kind of universe.
In the figure we label three phases: the bounce $B$, conventional $\Lambda$CDM cosmologies $\Lambda^\pm$ emerging in either temporal direction from the bounce, and a de~Sitter equilibrium phase $E$.
Let us investigate this picture.

In a typical cyclic cosmology, at some future time spatial slices reach a turnaround point where (at least in some conformal frame) they stop expanding and start contracting.
This might seem at odds with our model, where the equilibrium phase has no energy sources other than the cosmological constant; without something like spatial curvature or a time-dependent dark energy, an expanding universe will not stop and recontract.
But this intuition is somewhat misleading in our case.
Pure de~Sitter spacetime is maximally symmetric, and there is no coordinate-invariant notion of space either ``expanding" or "contracting" at any particular event; those notions only apply to slices in a particular foliation, not to individual spacetime points.
Slices with negative extrinsic curvature are said to be expanding, and those with positive extrinsic curvature are contracting.
So the resolution is that the natural foliation in which space appears to be expanding in the $\Lambda^+$ phase is not the one that starts contracting in the $\Lambda^-$ phase.

Rather, we can interpolate between these phases using the static-patch slicing of de~Sitter. The static metric takes the form
\be
  ds^2_\mathrm{dS} = -\cos^2(\chi) dt^2 + H_*^{-2}d\Omega_3^2,
  \label{dSstatic}
\ee
where $H_*$ is the (constant) Hubble parameter, $d\Omega_3^2 = d\chi^2 + \sin^2\chi(d\theta^2 + \sin^2\theta d\phi^2)$ is the metric on the round three-sphere, and the radial coordinate $\chi$ goes from $0$ to $\pi/2$, so the metric only covers an open half of the sphere.
These coordinates only apply to the causal diamond for an observer at the pole ($\chi = 0$), but that is appropriate since the entire universe in our model is represented by the horizon volume of one observer.
In this form, the radial dependence in $g_{00}$ means the metric is not Robertson-Walker, but constant-$t$ slices have zero extrinsic curvature (indeed, the geometry is reflection symmetric around any such surface).
It is therefore possible to match the geometry prior to any such slice to a time-reversed version to the future of the slice.

We therefore suggest a metric ansatz of the form
\be
 ds^2=-f^2(t, \chi) dt^2 + a^2(t)d\Omega_3^2.
\ee
In the equilibrium phase $E$, we should have $f(t,\chi) \approx \cos\chi$ and $a(t) \approx$~constant, recovering the de~Sitter metric (\ref{dSstatic}).
In the $\Lambda^-/B/\Lambda^+$ phases, we should have $f(t,\chi)\approx 1$ and the scale factor $a(t)$ describing the appropriate contraction/bounce/expansion behavior.

The conformal diagram for standard de~Sitter space is a square, with horizontal lines representing spatial three-spheres that expand without bound toward the past and future boundaries, and the left and right lines in the diagram representing poles of the spheres.
A universe that starts with a Big Bang singularity and asymptotes to de~Sitter would be a rectangle resembling the top half of that square, with the bottom boundary being the singularity.
In our diagram we instead have drawn a null boundary for these regions.
These are not literally boundaries to spacetime (in particular, they are not null infinity $\mathcal{I}^\pm$), but rather the boundary of the causal diamond corresponding to an ``observer'' at the pole of the sphere on the leftmost boundary of the diagram.
This is consistent with the idea, following Banks and Fischler \cite{banks+fischler2001}, that the finite dimensionality of Hilbert space is reconciled with the eternal expansion of classical de~Sitter spacetime by arguing that there are not \emph{independent} degrees of freedom characterizing what happens outside the cosmological horizon; rather, different observers describe (within the same finite-dimensional $\hs$) different causal diamonds, all related by unitary transformations of each other.

The universe in this picture does not undergo a sudden phase transition when it evolves from the equilibrium phase $E$ to $\Lambda^-$.
Instead, it is simply the evolution from $\Lambda^+$ to $E$ -- where the only remnant energy density comes from ever-redshifting photons that eventually equilibrate with the quantum de~Sitter vacuum -- played backward in time.
That is, it appears as if the vacuum gradually ``excites" to include a gas of ultra-long-wavelength photons of increasing density, which ultimately define a foliation of constant-density surfaces that continue to contract.
The photons form white holes that expel particles and larger objects, which eventually form (time-reversed) galaxies and stars, which then ultimately smooth out into a nearly-uniform dense plasma.
This all seems thermodynamically unusual, but is explained by the existence of an effective Future Hypothesis of low entropy at $B$, rather than the traditional Past Hypothesis (in our chosen global time coordinate).

Turning to the bounce phase $B$, standard wisdom, following the Penrose-Hawking singularity theorems \cite{hawking1970singularities}, holds that replacing the Big Bang with a bounce requires at the very least violation of the Null Energy Condition.
Various ways around this problem include the inclusion of exotic matter, branes and extra dimensions, or appeal to quantum effects \cite{gasperini1993pre,khoury2001ekpyrotic,Brown:2004cs,penrose2012basic,Battefeld:2014uga,Boyle:2018tzc}.
Alternatively, using a null boundary condition, Aguirre and Gratton derived a nonsingular bounce with a purely de~Sitter geometry \cite{aguirre2003inflation}.
In our picture, $B$ corresponds to the entropic minimum of the evolution of $|\psi\rangle$, $S_\mathrm{QB} = 0$.
On the diagram we have represented it as a spacelike surface, as it would be in conventional Big Bang cosmologies.
It is plausible that there is some semiclassical spacetime interpretation, but we would expect the system to be in a regime where quantum-gravity effects are important, and we shouldn't be too insistent on the universe obeying the classical Einstein equation at that moment.

Moving in either direction of time away from the bounce, we envision a conventional $\Lambda$CDM cosmology, labeled $\Lambda^\pm$ on the diagram.
The difference between crunch and bang is entirely one of convention; on both sides, entropy increases in the direction of time pointing away from the crunch, and observers living on either side would label the direction toward the bounce as ``the past." 

Although we have described the equilibrium phase $E$ as approximately de~Sitter, there are reasons to be cautious about such an identification.
Because the system is in equilibrium, there is no dynamical decoherence process that branches the state into a set of pointer states that represent distinct semiclassical geometries.
Indeed, there are no observers or test particles moving on an emergent metric.
There is only (aside from possible rare fluctuations) the equilibrium quantum state itself.
What were distinguishable branches of the wave function in the non-equilibrium phase are now macroscopically indistinguishable; the branches have essentially recohered into a single state, not because the environment states have lined up, but because the ``system" part of the wave function is simply empty de~Sitter spacetime in every branch.
There is no longer entanglement between a conventional system and environment, which is the hallmark of standard decoherence.

\section{Discussion}

Physics has yet to settle on a completely successful theory of all the fundamental particles and forces of nature.
It is possible that, in order to formulate such a model, it will become necessary to look past quantum mechanics as it is currently understood.
But given the incredible empirical success of quantum theory, and the difficulties involved in modifying it (including no direct empirical evidence that such modifications are needed), it is interesting and important to carefully consider the implications of conventional unitary quantum mechanics.
This paper provides one illustration of how such considerations can be fruitful even in the absence of a well-defined map between the fundamental quantum state and our emergent semiclassical world.

We have investigated some cosmological consequences of purely unitary (Everettian) quantum mechanics, featuring Schr\"odinger evolution with a time-independent Hamiltonian in a fixed Hilbert space.
Since time is eternal in such models, there is an important distinction to be made between finite- and infinite-dimensional Hilbert spaces, and here we focused on the finite-dimensional case.
Given some coarse-graining into macrostates, which we used to define a quantum Boltzmann entropy $S_\mathrm{QB}$, such a system will spend most of its time fluctuating around an equilibrium state of nearly-maximum entropy.
With a generic Hamiltonian (as characterized by its spectrum of energy eigenvalues), the state vector will evolve ergodically on a torus of phases, and exhibit dynamical fluctuations characterized by statistics of the form $P(\Delta S) \sim e^{-\Delta S}$.
This leads to the standard Boltzmann Brain problem, rendering such scenarios phenomenologically unacceptable.

In this paper we have identified a loophole in the reasoning of the previous paragraph, enabling us to construct a scenario with a pronounced thermodynamic arrow of time in the vicinity of a set of exactly periodic ``bounces."
We do not claim that our model is in any way generic; quite the opposite.
We invoke non-generic features in two ways.
First, we impose that the shifted energy eigenvalues be commensurable, i.e. $\widetilde{E}_k=p_k \omega_*$ for some integers $p_k$ and a fixed energy scale $\omega_*$.
This renders the evolution non-ergodic, and produces exactly periodic evolution rather than mere Poincar\'e recurrence.
Second, we choose an initial condition such that $S_\mathrm{QB}$ is at or near its minimal value.
This gives an entropy curve that spends most of its time near maximality, with rare periodic excursions that we interpret as a crunch/bounce/bang transition in spacetime.
Thermodynamic entropy in individual branches increases in both directions of time away from the minimum of $S_\mathrm{QB}$, providing a phenomenologically successful arrow of time.
On top of that, since the period of the evolution can be much less than the recurrence time, most observers in such a universe will arise naturally in the course of evolution from lower to higher entropy, rather than being fluctuations from equilibrium.
Finally, although it is not quite an example of fine-tuning, we do need to require that the fiducial energy scale $\omega_*$ be large enough (although still ridiculously small) to maintain compatibility with the bounds (\ref{Tbound}) on the period in order to avoid BB domination.

An obvious thing to ask is why the spectrum should feature commensurable eigenvalues, and why the initial condition should be dramatically low entropy in an appropriate coarse-graining.
We have no idea. For now we are confining ourselves to the more modest goal of verifying that this kind of scenario can possibly work. Nonetheless, pursuing a deeper explanation of this sort of fine-tuning is worthwhile when developing a more complete theory, though it may turn out that there is none. In that case, asking about fine-tuning would be equivalent to asking why the laws of physics take one form rather than another.

A (hopefully) more tractable problem is to better establish the emergence map from branches of the quantum state to an appropriate configuration of semiclassical quantities in a universe with a positive cosmological constant.
From the work on establishing how locality can emerge from the spectrum \cite{Cotler:2017abq}, it is suggested that a generic Hamiltonian has no local factorization into subsystems, and that when such a factorization exists, it is essentially unique.
(Although see \cite{Stoica:2021rqi} for a contrary view, and \cite{Soulas:2025kdn} for an argument that one also needs a specification of a state [which presumably our scenario would satisfy].)
It might seem unlikely that a spectrum of commensurable eigenvalues would also have the right properties to allow for emergent local spacetime, but it's reasonable to expect that approximate locality is broadly achievable \cite{loizeau2025quantum}, and presumably that would be empirically acceptable.

Even without a complete picture of how spacetime emerges from the evolving quantum state, it would be interesting to consider issues such as the homogeneity and isotropy of the universe, and the nature of density fluctuations, in this kind of framework.
Small steps have been made in this direction, thinking of cosmological evolution in terms of tensor networks \cite{Beny:2011vh,Czech:2015kbp,SinaiKunkolienkar:2016lgg,Bao:2017qmt} and quantum circuits \cite{Bao:2017iye}.
In the latter work, it was pointed out that the ``first entangled qubit" should have been created not much more than 70 $e$-folds before the end of inflation, raising the prospect of some kind of information persisting through the bounce phase.

The idea of a fixed finite-dimensional Hilbert space has implications for how we should think about fine-tuning problems in physics.
For example, as stressed by Banks and Fischler \cite{Banks:2018jqo}, in a fixed finite-dimensional Hilbert space it is not appropriate to think of the value of the cosmological constant as simply a coupling constant in a low-energy effective field theory; rather, it appears as an input parameter to the theory, related to the dimensionality of Hilbert space by $\Lambda \approx \pi/[G\log(\Dim\hs)]$.
This doesn't solve the cosmological constant problem, but turns it into an entirely different question -- why is the dimensionality of Hilbert space so large?

One might wonder whether the phenomenological acceptability of this model could provide support for finitism \cite{sep-geometry-finitism} or even ultrafinitism \cite{Cherubin2011AVS}, approaches to the foundations of mathematics that do away with infinity entirely.
The original motivation for considering commensurable eigenvalues was to construct a quantum model that could be consistently modified into a truly discrete theory with a finite number of elements \cite{Carroll:2023mzl}.
Ordinary quantum mechanics does not qualify, even in finite-dimensional Hilbert spaces, since both time and Hilbert space itself are continuum objects.
It is at least conceivable that a scenario like that presented here would enable a phenomenologically acceptable version of finitist physics that could be compatible with the experimental support for traditional quantum theory, with interesting consequences for the foundations of mathematics.

Finally, there are doubtless philosophical implications to be drawn from the idea that the universe is precisely periodic in time. Exact $t \rightarrow t + T_*$ symmetry over eternity leads to an infinite number of identical copies of the universe in its expansion and contraction phases (though the expansion and contraction phases are not identical to each other in the absence of exact $t \rightarrow -t$ symmetry).
As Nietzsche envisioned \cite{nietzsche1974gay}, everything you do has already been done infinitely many times in the past, and will be done infinitely many times in the future. 
We leave the moral implications of this view for readers to contemplate.
However, while we have presented our model as an example of periodic evolution in an eternal universe, the fact that the evolution is precisely periodic (not merely recurrent) opens the door to thinking of time as only lasting for a finite interval, simply by quotienting out the evolution by $T_*$.
Choosing to consider time in this model as defined on a circle rather than on a line is somewhat a matter of taste. 

Henri Poincar\'e, reflecting in 1893 on the relevance of his recurrence theorem to statistical mechanics,  mused that ``According to this theory, to see heat pass from a cold body to a warm one, it will not be necessary to have the acute vision, the intelligence, and the dexterity of Maxwell’s demon; it will suffice to have a little patience" \cite{poincare2003mechanism}.
With a little fine-tuning, he may turn out to be correct.

\section*{Acknowledgments}
This work was supported in part by a grant from the Wendell-Ousterhout Foundation.

\bibliographystyle{utphys}
\bibliography{cyclic_QM}
\end{document}